\documentclass[a4paper, amsfonts, amssymb, amsmath, reprint, showkeys, nofootinbib, twoside]{revtex4-1}

\usepackage[english]{babel}
\usepackage[utf8]{inputenc}
\usepackage[squaren,Gray,cdot]{SIunits}
\usepackage{graphicx}

\begin{document}
\title{High-magnification shadowgraphy for the study of drop breakup\\ in a high-speed gas flow}

\author{Luc Biasiori-Poulanges}
\author{Hazem El-Rabii}
\email[Correspondence email address: ]{hazem.elrabii@cnrs.pprime.fr}
\affiliation{Institut Pprime, CNRS (UPR 3346), 1 avenue Clément Ader, 86961 Futuroscope, France}

\begin{abstract}
Direct observation of the droplet breakup process in high-speed gas flows is a critical challenge that needs to be addressed to elucidate the physical mechanisms underlying the fragmentation phenomenon. Here, we present a high-magnification and high-speed shadowdograph technique that allows the visualization of this process over its whole evolution and resolves detailed features of the breakup zone. The developed experimental method uses a high-speed camera equipped with a long-distance microscope. The backlight illumination source is provided by the laser-induced fluorescence of a dye solution that delivers short pulses at a high-repetition rate. Artefacts resulting from the laser coherence are therefore reduced. 
\end{abstract}

\maketitle

Aerodynamic fragmentation of drops plays a key role in a rich variety of technical issues \cite{allison2016quantitative,bolleddula2010impact,eckhoff2016explosion}. Accordingly, this process has been extensively studied over past decades through an impressive number of investigations using a series of optical techniques \cite{dai2001temporal, kulkarni2012secondary,guildenbecher2009secondary,kim2012breakup,rajamanickam2017dynamics,gao2013quantitative, soni2019deformation}. Despite this, however, the mechanisms of fragmentation and their connections to the resulting size droplet distribution are far from being understood. The underlying reason likely is that these mechanisms involve multiphase flows whose dynamics exhibit extreme disparity between time/space scales, making their direct observations quite challenging. 

\begin{figure}[htbp]
	\begin{center}
		\centering
		\includegraphics[width=.9\columnwidth]{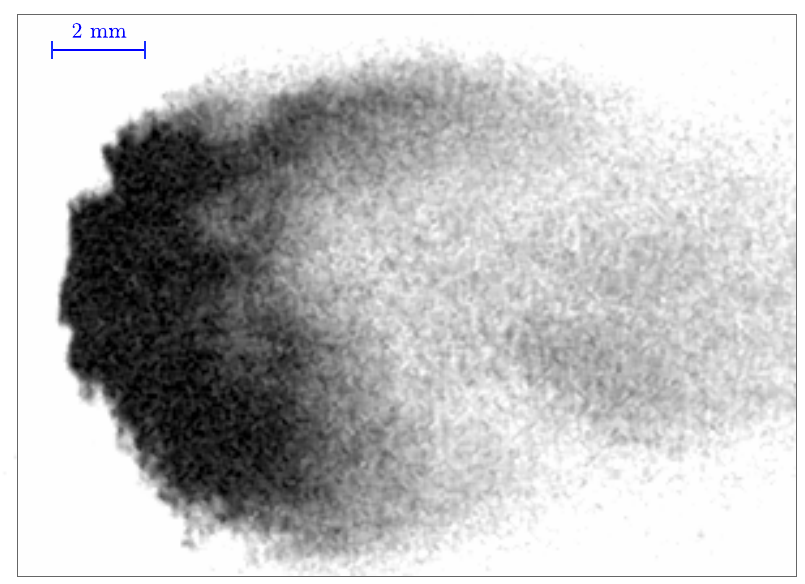}
		\caption{Conventional shadowgraphy (integration time: 350 \nano\second, resolution: 17 pixel$\cdot$\milli\reciprocal\meter) at $\tau$=1.43, $\mathrm{We}$=1350.}
		\label{Figure1}
	\end{center}
\end{figure}

Visualization of drop fragmentation in gas flow has mostly been realized by the shadowgraph technique that has provided valuable information on the overall process (see, for instance, \cite{hanson1963shock, soni2019deformation} and references therein). Past studies have identified various breakup mechanisms and regimes ranging from vibrational to catastrophic breakup depending on the Weber number, $\mathrm{We}$ (ratio of inertia to capillary forces). The limited spatial and temporal resolution of the shadowgraph technique, however, has restricted its successful applicability to the different bag-breakup modes, that is, for $\mathrm{We}<100$. With increasing $\mathrm{We}$, a thin sheet at the periphery of deforming drop appears and breaks up continuously, producing a dense mist surrounding the drop (Fig.~\ref{Figure1}). The onset of this process marks the transition to shear-type breakup regimes, which are reached for $\mathrm{We}>1000$. 
In such conditions, the shadowgraph technique has failed in capturing features related to the primary and secondary atomization process (e.g., ligaments, secondary droplets), producing image quality inadequate to any conclusive interpretation. 
At present, high-spatial resolution can be reached, but at the cost of a low-temporal resolution. In that case, images recorded from repeated experiments (performed under the very same conditions), with progressing time delay, are stitched together into one sequence. This assumes that the breakup event is reproducible such that all the frames correspond to the exact same initial conditions. While the mean behavior of the breakup process seems reproducible, micro-scale processes (e.g., instabilities) vary strongly from one breakup event to the other. Time-resolved imaging of these processes required then that their dynamics be recorded from a single breakup event. With the advent of ultra-high-speed cameras, it is now possible to acquire videos at frame rates and pixel resolutions that are compatible with high Weber number regimes. To resolve sub-millimetric spatial scales associated with the morphology evolution for these breakup regimes, however, a long-range microscope optical system is needed. This requires in turn a bright illumination source with a sufficiently short duration (sub-microsecond) to avoid motion blur and loss of resolution. Despite their brightness and short pulse durations, lasers are poorly suited for this purpose because their high spatial coherence introduces artefacts (e.g., speckle) that significantly degrade image quality. Laser driven light sources or light emitting diodes, on the other hand, produce images with limited resolution and signal-to-noise ratio.

In this Letter, we show that the fluorescence light of Rhodamine 6G dissolved in ethanol and excited by a 532\,\nano\meter~pulsed laser at high repetition rates (up to 80\,\kilo\hertz) provides a light source that is ideally suited for the considered application. Indeed, it has a definite advantage because short light pulses can be generated at a rate that matches the frame recording rates of high-speed cameras, while minimizing coherence. Combined with a high-magnification optical system the diagnostic proves successful in producing highly time- and space-resolved images of the drop aerobreakup process, for Weber number as high as 1350. Sequence of images so obtained gives a complete picture of breakup development, from its very outset (deformation and instabilities development) to the ultra-fine droplets ultimately generated after the secondary-breakup stage, unveiling detailed features of the evolving internal structure of the breakup zone. 
\begin{figure}[htbp]
	\begin{center}
		\centering
		\includegraphics[width=\columnwidth]{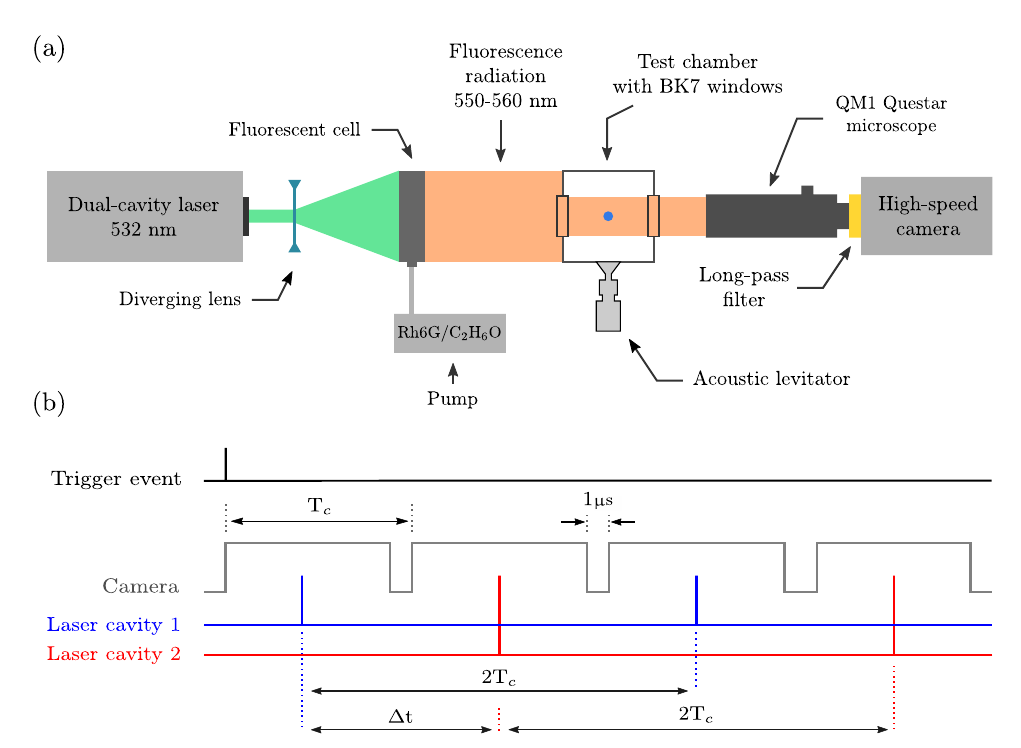}
		\caption{(a) Sketch of the experimental arrangement (axial view), (b) Timing diagram for camera exposure and laser pulses.}
		\label{Figure2}
	\end{center}
\end{figure}
	
Figure~\ref{Figure2}~(a) shows a schematic of the experimental set-up used in this study, which includes an air shock tube, a single‐axis acoustic levitator, and a high-magnification shadowgraphy system. 

The shock tube is employed to produce a transient flow of known velocity. It is composed of a driver section  separated from a driven section by a double-membrane section. The driven section comprises a test chamber of square section and circular cross-section pipes connected together by means of circular-to-square transitions. The test chamber is fitted with two oblongs BK7 windows, which are mounted opposite one another on its lateral sides. During the experiments, the water drop is held in a stable equilibrium at the center of the test chamber by the sound radiation pressure of an ultrasonic standing wave generated by the single-axis acoustic levitator. The levitation system comprises a Langevin-type transducer coupled to a mechanical amplifier with a concave radiating surface. The radiating surface is mounted flush with the inside bottom surface of the chamber. Opposite to it, the upper surface of the chamber acts as a reflector of the acoustic waves for standing wave generation.

The high-magnification shadowgraph imaging system includes a high-power laser coupled to a fluorescent cell, and a long distance microscope mounted on a high-speed camera. The laser is a dual oscillator/single head diode-pumped Nd:YAG laser (Mesa PIV, Continuum) operating at 532 \nano\meter. It produces short pulses with duration in the 120–180 \nano\second~range at repetition rates up to 40 \kilo\hertz~(for each oscillator), with an output average power of 60 \watt. The combination of the two oscillators enables the generation of pulse pairs with adjustable temporal delay down to 2 \micro\second. The laser beam is first expanded by a concave spherical lens and then fed into a fluorescent cell filled with a mixture of rhodamine 6G dye dissolved in absolute ethanol ($4.2\times 10^{-5}$ \mole\per\litre). To prevent photodegradation of the dye molecules, the solution is continuously flowed through the cell, using a pump circulation system, to insure a fast renewal of the dye molecules exposed to the laser. The resulting fluorescence light leaving the cell is incoherent and shifted from 532 \nano\meter~to 550-650 \nano\meter. The long distance microscope is a Maksutov–Cassegrain catadioptric microscope (QM1 Questar) with a working distance of 560 \milli\meter. The maximum optical resolution is 1.6 \micro\meter~at 560 \milli\meter~with a magnification up to 125:1. The depth of focus is approximately 0.6 \milli\meter. The microscope is mounted on a Fastcam Photron SA-Z with a 1024$\times$1024 pixel resolution and a maximum frame rate of 20 \kilo\hertz. The maximum frame rate of 480 \kilo\hertz~is obtained with a 512$\times$56 pixel resolution. An optical long-pass filter with a cutoff wavelength of 550 \nano\meter~is mounted between the far-field microscope and the camera to block remaining green laser light transmitted through the fluorescent cell. The high-speed camera and the dual-cavity laser are synchronized by a digital delay generator, according to the timing diagram shown in Fig.~\ref{Figure2}~(b). A typical sequence that records an image series proceeds as follows: First a pressure transducer used to detect the passage of the shock wave (at some specific location in the tube) delivers a trigger signal to the delay generator. The latter then initiates camera acquisition at a frame rate of $1/T_c$. The interframe time of an image sequence is 1 \micro\second. Within a short time period following the trigger event, the laser system generates two laser pulse trains with a pulse repetition frequency of $1/2 T_c$. The pulses in the first and second pulse trains are delayed in relation to each other by a predefined and varying time delay $\Delta t$. One frame out of two is recorded with the first pulse train, whereas the pulses of the second pulse train provide illumination for the other frames.  

To assess the capacity of the technique to provide new and detailed insights into the spatio-temporal dynamics of the fragmentation process, we performed experiments to record data of single drop breakup events for three diameters (0.8, 1.6 and 2.0\,\milli\meter) at a shock Mach number $M_\mathrm{s}=1.3$. The corresponding $\mathrm{We}$ are 540, 1080 and 1350, respectively. These conditions do not require to operate the optical diagnostic at its full potential. Therefore, the camera settings were adjusted to record two frames at 1024$\times$1024 pixel resolution with a separation of 8 \micro\second~at a frequency of 20 \kilo\hertz. Each cavity of the high-power laser ran at 10 \kilo\hertz~with an average output power of 30 \watt~and a pulse width of 174 \nano\second. The field of view was approximately 6.5$\times$6.5 \milli\meter\squared~and the measured spatial resolution of the imaging system was 6.5 \micro\meter~per pixel. 

\begin{figure}[htbp]
	\begin{center}
		\centering
		\includegraphics[trim={0cm 0.1cm 0cm 0.1cm},clip,scale=0.247]{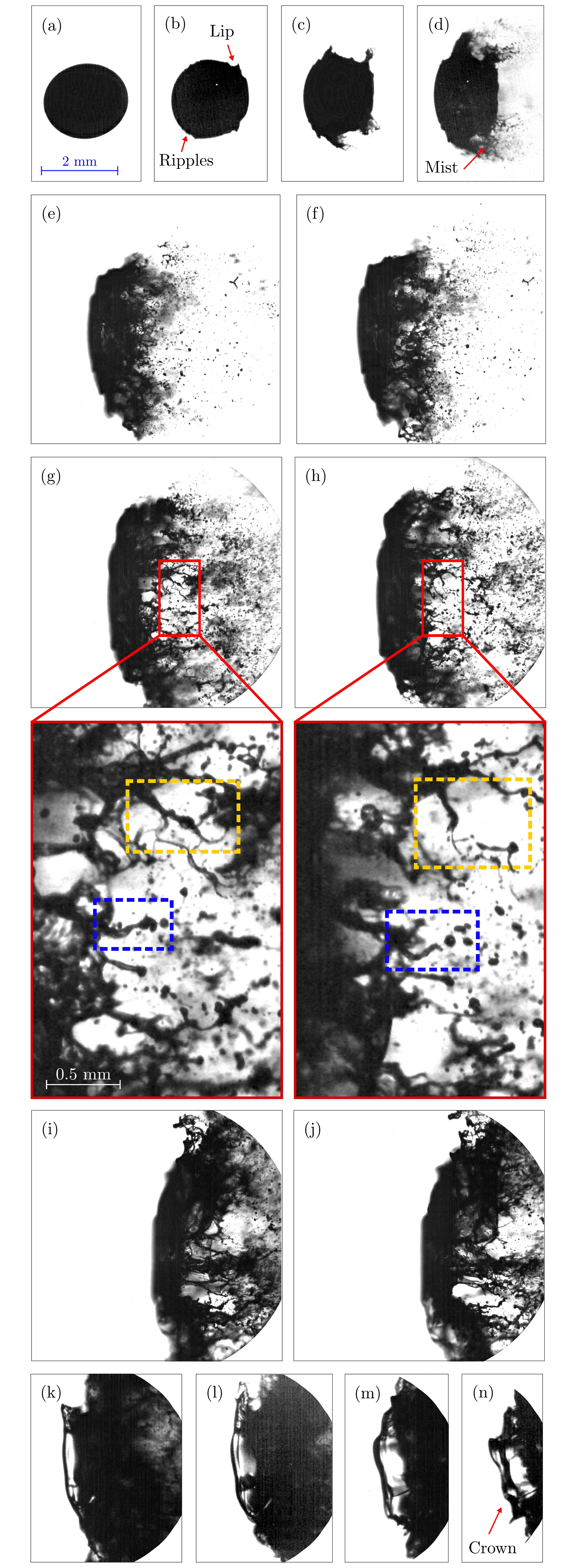}
		\caption{High-magnification shadowgraphy ($\mathrm{We}$=1350). Times $\tau$ are (a) 0.04, (b) 0.22, (c) 0.28, (d) 0.46, (e) 1.18, (f) 1.24, (g) 1.42, (h) 1.48, (i) 1.66, (j) 1.72, (k) 1.90, (l) 1.96, (m) 2.14, (n) 2.20.}
		\label{Figure3}
	\end{center}
\end{figure}

Here, time is measured in units of the characteristic time $\tau_c=d\sqrt{\theta}/V$, where $d$ is the initial drop diameter, $\theta$ is the drop-to-gas density ratio and $V$ is the initial relative velocity between the flow field and the drop using the post-shock conditions. The time origin corresponds to the shock-drop interaction event. Figure~\ref{Figure3} shows selected sequential images captured by the high-magnification shadowgraph system after a levitated drop of 2.0\,\milli\meter~in diameter was struck by the planar shock wave ($\mathrm{We}$=1350). The shock wave direction is from left to right. 

During the first moments ($\tau<0.1$), the drop keeps its original shape and seemingly remains undisturbed (Fig.~\ref{Figure3}~(a)). At the end of this period, the effects of the interaction are manifest by the appearance of a small lip on the top and bottom of the leeward face of the drop (Fig.~\ref{Figure3}~(b)). The formation of the lips is immediately followed by the development of ripples over the whole drop surface. As time elapses, the ripples and lips increase and bent in the flow direction producing a trail of mist under the destabilizing effect of the airstream. The drop simultaneously undergoes a continuous deformation, flattening in the streamwise direction with the windward face keeping a nearly spherical-cape shape while the leeward face is flattened into a planar surface. The shape of the drop evolves into a muffin-like one (Fig.~\ref{Figure3}~(c)-(d)), before taking on a crescent-like body shaped (Fig.~\ref{Figure3}~(e)). Up to this point, the deformation and breakup of the water drop is basically consistent with previous results found in the literature. It is important to note, that under the same experimental conditions, traditional shadowgraphy is inefficient in providing any reliable information on the subsequent evolution of the fragmentation process due to limitation in spatial resolution. This limitation is illustrated in Fig.~\ref{Figure1}, which displays an image corresponding to $\tau=1.43$ and obtained by a traditional shadowgraph technique under the same experimental conditions.  

On the contrary, it is seen in Fig.~\ref{Figure3} that the high-magnification arrangement effectively resolves features related to the primary and secondary atomization process. Several observations can be made from this figure. We note that most of the mist is first blown from the crescent horns in a conelike structure that becomes denser and longer. On the remaining part of the deformed drop surface, the formation of multiple streamwise ligaments connected to the liquid core is seen in Fig.~\ref{Figure3}~(f). Droplets and various liquid packets of different sizes resulting from the breakup of these ligaments can be observed downstream of the liquid core. Figure~\ref{Figure3}~(g) shows more clearly that the entire breakup zone contains an intact liquid core, followed by an interconnected ligaments zone, which is the transitional zone between the liquid core and the micro-droplet zone.  Zoomed-in selection of the ligament zone, indicated by the red frame, are displayed below Fig.~\ref{Figure3}~(g)-(h). The image reveals an intricate ligament-branching structure in which there are threads and dendritic forms of liquid, as well as ligaments alone or breaking up into one or more drops by the Plateau-Rayleigh instability or end pinching. Figures~\ref{Figure3}~(g)-(h) show pairs of images taken 8\,\micro\second~apart. For example, the blue-frame in images (g)-(h) capture the formation of a droplet at the free end of a ligament by end-pinching mechanism; the yellow-frame show the breakup of a ligament into two ligaments. One can also see that the threads and ligaments formed during this stage vary widely in their sizes, eventually producing thereby droplets of high polydispersity.  It is important to note that the high spatial resolution of the diagnostic allows accurate size determination of all these liquid fragments from images. At subsequent time (see images Fig.~\ref{Figure3}~(k)-(n)), we observe the emergence of  a water sheet expanding in the opposite direction of the flow. A careful examination of the images recorded at previous instants (not shown here) suggests that this water sheet originates from the stretching, in the counter-streamwise direction, of a protrusion structure that developed on the upstream surface of the flattened drop. The liquid sheet takes an asymmetrical crown shape. At later time ($\tau$ > 2.20), the images captured by the diagnostic show the formation of a water sheet in the shape of a spherical cap that appears to emerge from inside the crown. To the best of our knowledge, this is the first reported observation of such a morphological evolution for the drop fragmentation process.
\begin{figure}[htbp]
	\begin{center}
		\includegraphics[width=\columnwidth]{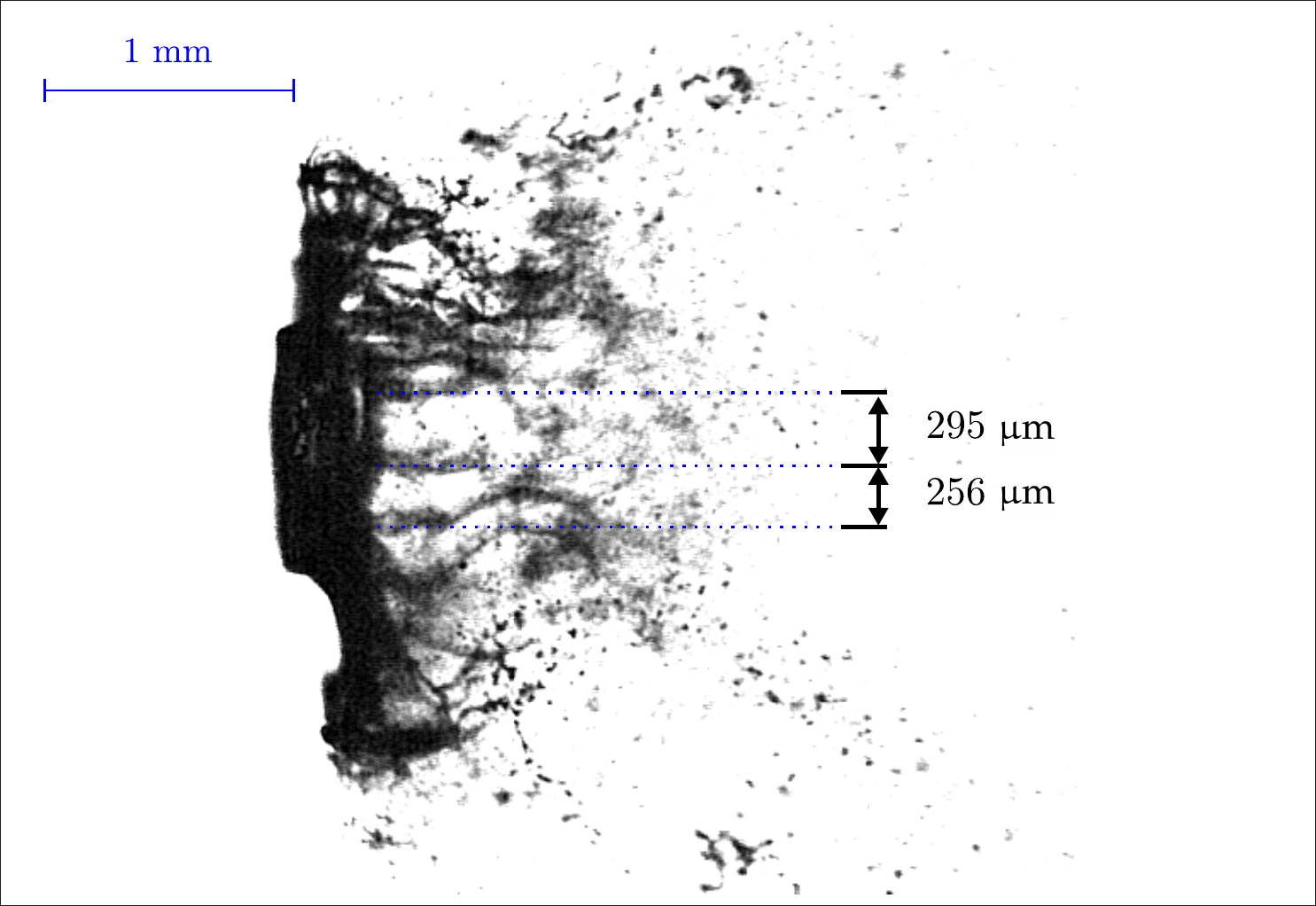}
		\caption{High-magnification shadowgraphy ($\mathrm{We}$=540) at $\tau$=0.95.}
		\label{Figure4}
	\end{center}
\end{figure}

Another striking example of features unveiled by the diagnostic is offered in Fig.~\ref{Figure4}. The image shows a snapshot (at $\tau=0.95$) of a water drop, with an initial diameter of 0.8\,\milli\meter, after its interaction with a Mach 1.3 shock wave. It is seen that ligaments formed during the fragmentation process emerge from the flattened droplet's edge at periodic transverse azimuthal positions. Such a periodic ligamentary structure is expected from numerical simulations \cite{jalaal2014transient,meng2018numerical} and has been tentatively interpreted as due to a combinaison of two types of instability. A shear instability first develops and generates axisymmetric surface waves on the droplet boundary. These waves induces a transient accelerations perpendicular to the gas/liquid interface which then trigger a transverse instability causing azimuthal modulations. These modulations grow in amplitude producing ligaments at the wave crests. It was suggested in \cite{jalaal2014transient} that the transverse instability responsible for the modulations is the Rayleigh-Taylor instability (RTI), but poor quantitative agreement between their numerical simulations and theory does not support such an assumption. It is however worth pointing out that, irrespective of mechanisms causing the azimuthal modulation, the computed modulation wavelength from numerical simulations must agree with the measured one. This point is confirmed by the measured modulation wavelength value obtained from Fig.~\ref{Figure4}, which is approximately $\lambda$=280\,\micro\meter. This corresponds to a dimensionless wavenumber (2$\pi$d/$\lambda$) of 18, in agreement with the range of values obtained from the numerical simulations reported in \cite{jalaal2014transient}.
\begin{figure}[htbp]
	\begin{center}
		\centering
		\includegraphics[width=\columnwidth]{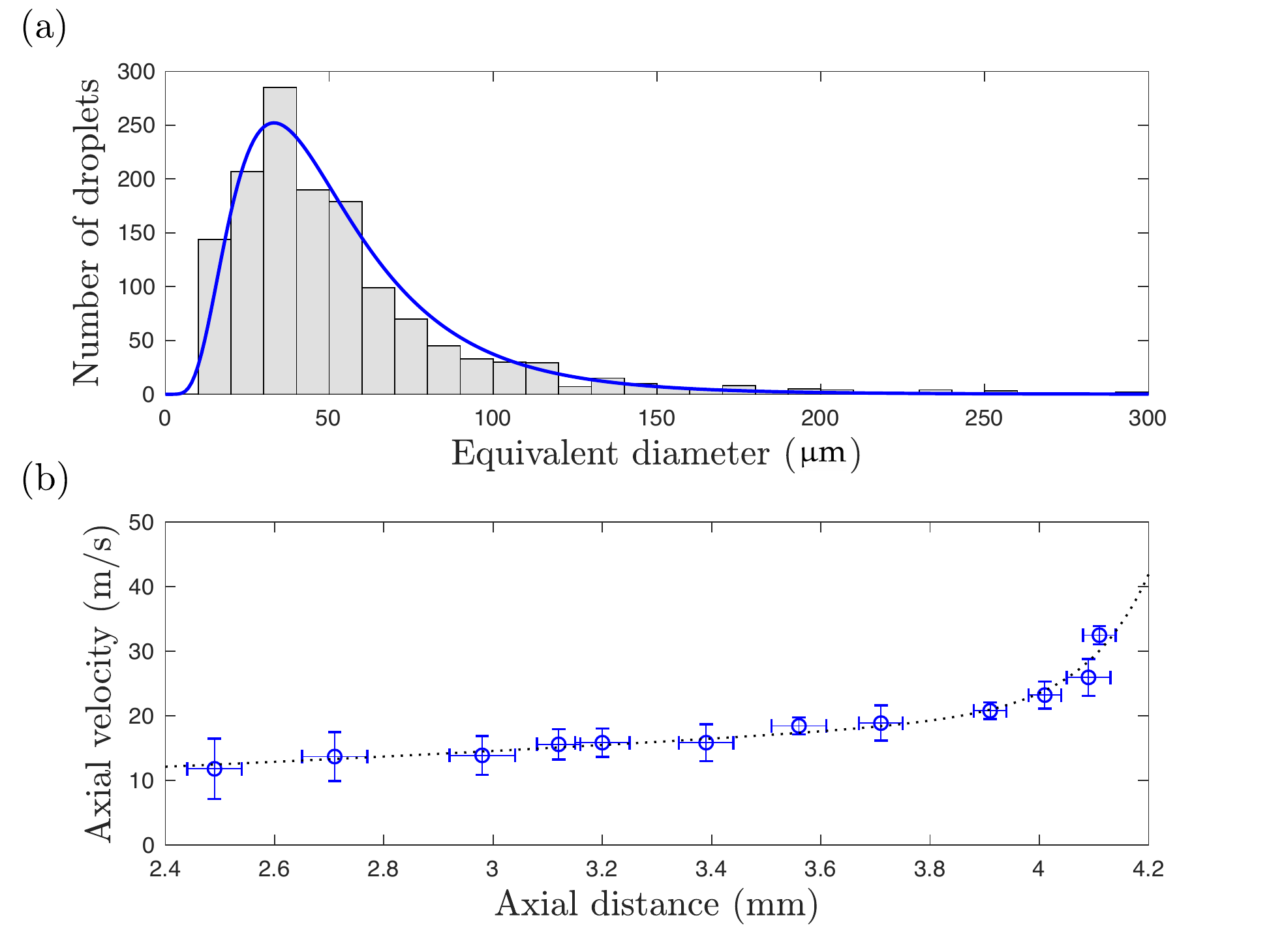}
		\caption{(a) Droplet size distribution (gray bins). The blue line is a log-normal fit. (b) Axial velocities against axial distance from an arbitrary reference point (each blue marker corresponds to a different drop).}
		\label{Figure5}
	\end{center}
\end{figure}

Additional quantitative data can be extracted from the recorded video frames. Specifically, measurement of secondary droplet characteristics, such as size distributions and axial velocities, can be obtained through image processing techniques. Figure~\ref{Figure5}(a) is an example of a typical histogram of secondary droplet size distribution constructed from a sample of 1600 droplets that results from the interaction of a 1.6\,\milli\meter~water drop with a Mach 1.3 shock wave ($\tau=0.88$). The size measurements (not corrected for droplet overlapping error) were inferred from a single shadowgraph image using a  classical processing procedure: background substraction, contrast enhancement, and contour detection. The reported size values are the equivalent diameters defined as the diameter of a circle with the same projected area as the droplet. The size distribution given here shows that the imaging system used has a spatial resolution able to resolve secondary droplet dimensions produced in breakup regimes with $\mathrm{We}$ as high as 1000. Due to the high acquisition rate of the camera at a pixel resolution compatible with secondary droplet sizing, the diagnostic is capable of fully resolving rapid changes in droplet positions. Therefore, by tracking the same droplet in two consecutive images separated by a short (controllable) time interval, the diagnostic provides axial velocity measurements of secondary droplets (see Fig.\,\ref{Figure5}(b)). 

Unfortunately, we have not been able to find measurements, made under the same (or close) experimental conditions, that can be compared to ours. We can, however, note that the drop size distribution reported in \cite{gao2013quantitative}, although obtained in the bag breakup regime, is quite close to ours.

In conclusion, our experiments show that the shadowgraph technique presented here allows the visualization of the fragmentation process with high spatio-temporal resolution (for $\mathrm{We}$ up to $10^3$) over its whole evolution, capturing detailed features of the breakup zone and new morphological evolutions.\\

\noindent {\textbf{Funding.}}\,R\'egion Nouvelle-Aquitaine (SEIGLE project 2017-1R50115; CPER FEDER project). \\

\noindent {\textbf{Acknowledgment.}}\,The authors gratefully acknowledge the technical assistance of Jean-Carl Rousseau and Alain Claverie.\\

\noindent {\textbf{Disclosures.}}\,The authors declare no conflicts of interest.

\bibliography{references}

\end{document}